\documentclass[twocolumn,showpacs,floatfix,aps,nofootinbib]{revtex4}
\usepackage[dvips]{graphics,color}
\usepackage{epsfig}\usepackage{float}
\RequirePackage{amssymb}
\usepackage{makeidx}
\usepackage[brazil]{babel}
\usepackage[latin1]{inputenc}
\usepackage{graphicx}
\usepackage{indentfirst}
\usepackage{color}
\newcommand{\be}{\begin{equation}}
\newcommand{\ee}{\end{equation}}

\newcommand{\ba}{\begin{eqnarray}}
\newcommand{\ea}{\end{eqnarray}}

\begin{document}

\title{Notes on the Two-brane Model with Variable Tension}

\author{M. C. B. Abdalla$^{1}$}
\email{mabdalla@ift.unesp.br}
\author{J. M. Hoff da Silva$^{1}$}
\email{hoff@ift.unesp.br}
\author{R. da Rocha$^{2}$}
\email{roldao.rocha@ufabc.edu.br}

\affiliation{1. Instituto de F\'{\i}sica Te\'orica, Universidade
Estadual Paulista, R. Dr. Bento Teobaldo Ferraz 271 - Bl. II -
Barra Funda 01140-070 S\~ao Paulo, SP, Brazil}

\affiliation{2. Centro de Matem\'atica, Computaç\~ao e Cogniç\~ao,
Universidade Federal do ABC, 09210-170, Santo Andr\'e, SP, Brazil}


\begin{abstract}
Motivated by possible extensions of the braneworld models with two
branes, we investigate some consequences of a variable brane
tension using the well established results on consistency
conditions. By a slight modification of the usual stress-tensor
used in order to derive the braneworld sum rules, we find out some
important constraints obeyed by time dependent brane tensions. In
particular it is shown that the tensions of two Randall-Sundrum
like branes obeying, at the same time, an Eötvös law, aggravate
the fine tuning problem. Also, it is shown that if the hidden
brane tension obeys an Eötvös law, then the visible brane has a
mixed behavior allowing a bouncing-like period at early times
while it is dominated by an Eötvös law nowadays. To finalize, we
discuss some qualitative characteristics which may arise in the
scope of dynamical brane tensions, as anisotropic background and
branons production.
\end{abstract}
\maketitle

\section{Introduction}

In the last years the amount of works dealing with the possibility
of a universe with extra dimensions is widely increasing. It is,
in part, due to formal progress in string theory \cite{HW}, but
also (and perhaps mainly) to the possibility of the hierarchy
problem explanation \cite{PREC2}. Moreover, the scenarios
investigated by Randall and Sundrum (RS) in \cite{RSI,RSII} allow
the existence of a large extra dimension in a non-factorizable
background geometry, where an exponential warp factor intervenes
in the Higgs mechanism on the brane, making any hierarchy
unnecessary. The basic setup of the RS model \cite{RSI} is
composed by two mirror domain walls (the 3-branes) placed at the
extremity of an extra transverse dimension, an
$S^{1}/\mathbb{Z}_{2}$ orbifold. Our universe is described by one
of those branes (the so-called visible brane) and the metric is
established due to a quite important (and necessary) constraint
between the branes tensions: they must be equal and opposite. That
last characteristic is the so-called RS model fine tuning.

Recently, in the context of braneworld scenarios \cite{RSII}, the
hypothesis of the universe being better described by a variable
brane tension \cite{LAS} has been raised. In fact, keeping in mind
the cosmological evolution of the universe, a variable brane
tension appears as a necessity. Such a possibility, however,
should be explored with great care. To solve the Einstein's
equations for a variable tension brane in a complete scenario,
with all the required constraints for a braneworld model, is far
from trivial. For an illustrative example let us consider a
RS-type model with the hypothesis of variable tension. Following
the notation and conventions used in \cite{RSI}, one sees that the
variation on the tension is achieved by $V_{vis}$ and $V_{hid}$,
the vacuum energy of the visible and the hidden brane,
non-constants. Consistency is achieved if the term $\sigma(\phi)$
in the warp factor is replaced by $\sigma(x^{\mu},\phi)$, where
$\phi$ is the extra dimension and $x^{\mu}$ are the non-compact
brane dimensions. Let us restrict the analysis to the case in
which there is only time dependence on the warp factor, i. e.
$\sigma(t,\phi)$ since, in the light of \cite{LAS}, it is the
physical-motivated situation. In such case, when the Einstein's
equations are computed, starting from a diagonal metric {\it
ansatz} $(G_{\mu\nu})$, a non-diagonal component of the Einstein
tensor appears in the form \ba
\sqrt{-G}\frac{\partial^{2}\sigma}{\partial t\partial
\phi}&=&\left.-\frac{1}{12M^{3}}[\Lambda
\sqrt{-G}G_{04}+V_{vis}\sqrt{-g^{vis}}g_{04}^{vis}\right.\nonumber\\&\times&
\left.\delta(\phi-\pi)+V_{hid}\sqrt{-g^{hid}}g_{04}^{hid}\delta(\phi)]
,\right.\label{1} \ea and since the right hand side of the above
equation is zero, one arrives at $\partial^{2}\sigma/\partial
t\partial \phi=0$, which invalidates the premise. Besides,
physical arguments point into the same direction: $\sigma$ is
present in the warp factor and, consequently, it enters in the
rescaling of masses in the Higgs mechanism. Therefore it cannot be
time dependent.

The crucial point is that once the idea of a constant tension is
not taken into account in the two-brane model, the line element
may be no longer diagonal, since it may also lead to an
anisotropic background. In such a case the brane tension cannot be
understood as the brane vacuum energy, since the brane is not a
boost-invariant isotropic brane anymore \cite{tenso}. It is
possible to show, however, that starting from an off-diagonal
five-dimensional metric, after a diagonalization by an anholonomic
frame, the hierarchy problem can be solved \cite{OD}.

We should emphasize, however, that the variable braneworld tension
models arising from a generalization of \cite{RSII} do not lead to
an anisotropic background. In the case of an Eötvös brane tension,
like the one studied in reference \cite{LAS}, it is shown a
complete solution highly compatible to the observable symmetries.

In the presence of a complete solution obtained from first
principles, it is important to support the model by respecting
general consistence conditions obtained from Einstein's equations,
when applied to the braneworld scenario. The consistency
conditions are even more essential in the absence of a complete
solution. Such conditions were first developed in the context of
five-dimensional braneworlds \cite{GKL} and then extended to
arbitrary dimensions \cite{Leblond}. In particular, those works
corroborate the necessity of equal brane tensions --- of opposite
signs --- in RS model \cite{RSI} in order to describe a nearly
flat universe.

The main aim of this paper is to look at the consistency
conditions of RS-like braneworlds in five dimensions taking into
account variable tensions in both mirror branes. It is possible to
implement a variable brane tension into the general stress-tensor
used to develop the braneworld sum rules. We are specially
concerned about Eötvös branes, i. e. branes whose tensions are
given by the phenomenological law\footnote{The subindex of $T_{3}$
denotes a 3-brane.} \cite{eotvos} \be
T_{3}=\lambda(T_{c}-T)\label{doido},\ee where $\lambda$ is a
constant and $T_{c}$ the critical temperature above what any brane
exists. The usual interpretation is given as follows: the tension
$T_{3}$ is an intrinsic characteristic of the brane and it
increases as the temperature continuously decreases with time.
Therefore, as the universe expands the temperature decreases and
the brane becomes more rigid. The case encoded in the equation
(\ref{doido}) is the physical one analyzed in reference \cite{LAS}
for the one-brane model. Here we study some possible
configurations for a time dependent brane tension. In particular
we show that two mirror branes endowed with Eötvös tensions at the
same time are {\it not allowed}, except if the branes tensions
obey a dynamical fine tuning, i. e., if the branes tensions are
equal and opposite, time-dependent. It is, in some sense, an
aggravation of the fine tuning problem, indicating that both
branes should at least be part of a more general (unknown) global
dynamics.

Going one step further, it is shown that, if the hidden brane gets
an Eötvös tension, then the visible brane tension is dominated by
an exponential damping factor at early times, showing an
Eötvös-like behavior nowadays. The damping era, in which the brane
tension decreases, is finished at a specific time, say $\bar{t}$,
which denotes the inflection point of the tension variation.
Keeping the usual interpretation of variable tensions, the present
time (of increasing tension) can be associated with an expanding
universe, while the damping era (before $\bar{t}$) may be
understood as a bouncing-like behavior of the universe.

We emphasize that all this paper content is based upon toy models,
since a complete scenario for a dynamical tension must be obtained
from first principles. However, it is still interesting to study
those specific cases. The impossibility of both branes present
Eötvös tensions at the same time, and a bouncing-like behavior of
the visible brane, are quite important characteristics which can
serve, in some sense, as guidelines for the study of more
elaborated variable braneworld tension models.

We remark that in the string theory framework, the possibility of
variable tensions was analyzed in several contexts. For example,
in \cite{ST1} it was implemented as an integration constant while
in \cite{ST2} as a dynamical variable. Apart of that, some
stabilization mechanisms as well as supersymmetric branes in
singular spaces also used variable brane tensions, but in those
cases the tension depends on a bulk scalar field in order to
stabilize the distance between the branes \cite{ST3}, and also on
the superpotential \cite{ST4}, respectively.

Here, we shall keep the scope of the analysis only about
braneworld models. In fact, we are not concerned with the
mechanism under which the tension becomes variable, instead we are
looking for possible dynamics it obeys. As a first approximation,
we consider only time variable tensions. In this way, the Eötvös
law (\ref{doido}) should be replaced by \be
T_{3}^{(i)}=\pm\lambda^{(i)}t+\beta^{(i)},\label{mdoido}\ee where
$\lambda^{(i)}$ is a positive constant and $\beta^{(i)}$ a
constant representing the lower value for the brane tension. The
upper index $(i)$ denotes a specific brane and runs in the range
$i=vis,hid$, the visible and the hidden brane. In what follows we
absorb the constant $\beta^{(i)}$ into the definition of
$T_{3}^{(i)}$, just for convenience,  in such a way that
eq.(\ref{mdoido}) can be rewritten just as
$T_{3}^{(i)}=\pm\lambda^{(i)}t$. Note that the standard
interpretation still holds for the universe evolution: for a
positive tension, for instance, as the universe expands --- and
cools --- the tension increases. We just change the order
parameter from temperature to time. In fact, it can be
accomplished just in {\it very strict cases} in which the relation
between time and temperature is linear. We assume that it is the
case for an Eötvös tension. As we will see in the next Section
(C), this assumption does not preclude the possibility of non
linear effects in the tension.

This paper is organized in the following way: in the next Section
we derive the consistency conditions for braneworld in five
dimensions in the scope of variable tensions. After that, we study
the possible time variations in three specific cases: both branes
endowed with Eötvös tensions, both branes respecting the obtained
sum rules independently, and the last case when we fix the hidden
brane tension as an Eötvös one. We finalize pointing out some
important effects to be explored in the framework of variable
brane tension models.

\section{Allowed dynamic tensions}

In this Section we shall reobtain, for book-keeping purposes, the
basic results about consistency conditions in five dimensions, in
close analogy to the results in \cite{GKL,Leblond}. This
preliminary setup allows the future analysis of the possible
dynamics for the branes tensions.

We analyze a $D$-dimensional bulk spacetime endowed with a
non-factorizable geometry, characterized by a metric given by \ba
ds^{2}&=&\left.G_{AB}dX^{A}dX^{B}\right.\nonumber\\&=&
\left.W^{2}(r)g_{\alpha\beta}dx^{\alpha}dx^{\beta}+g_{ab}(r)dr^{a}dr^{b}\label{2},\right.
\ea where $W^{2}(r)$ is the warp factor, $X^{A}$ denotes the
coordinates of the full $D$-dimensional spacetime, $x^{\alpha}$
stands for the $(p+1)$ non-compact coordinates of the spacetime,
and $r^{a}$ labels the $(D-p-1)$ directions in the internal
compact space. The $D$-dimensional Ricci tensor can be related to
its lower dimensional partners by \cite{GKL} \begin{eqnarray}
R_{\mu\nu}&=&\bar{R}_{\mu\nu}-\frac{g_{\mu\nu}}{(p+1)W^{p-1}}\nabla^{2}W^{p+1},\label{2}
\\
R_{ab}&=&\tilde{R}_{ab}-\frac{p+1}{W}\nabla_{a}\nabla_{b}W,\label{3}\end{eqnarray}
where $\tilde{R}_{ab}$, $\nabla_{a}$ and $\nabla^{2}$ are
respectively the Ricci tensor, the covariant derivative and the
Laplacian operator constructed by means of  the internal space
metric $g_{ab}$. $\bar{R}_{\mu\nu}$ is the Ricci tensor derived
from $g_{\mu\nu}$. Denoting the three curvature scalars by
$R=G^{AB}R_{AB}$, $\bar{R}=g^{\mu\nu}\bar{R}_{\mu\nu}$ and
$\tilde{R}=g^{ab}\tilde{R}_{ab}$ we have, from equations (\ref{2})
and (\ref{3}), \ba
\frac{1}{p+1}\Big(W^{-2}\bar{R}-R^{\mu}_{\mu}\Big)=pW^{-2}\nabla
W\cdot\nabla W+W^{-1}\nabla^{2}W \label{4}\ea and \be
\frac{1}{p+1}\Big(\tilde{R}-R_{a}^{a}\Big)=W^{-1}\nabla^{2}W,\label{5}
\ee where $R^{\mu}_{\mu}\equiv W^{-2}g^{\mu\nu}R_{\mu\nu}$ and
$R^{a}_{a}\equiv g^{ab}R_{ab}$. Note that with such notation we
have $R=R^{\mu}_{\mu}+R^{a}_{a}$. It can be verified that for an
arbitrary constant $\xi$ the following identity holds  \be
\frac{\nabla \cdot (W^{\xi}\nabla W)}{W^{\xi+1}}=\xi W^{-2}\nabla
W\cdot \nabla W+W^{-1}\nabla^{2}W \label{6}.\ee Combining the
equation above with eqs. (\ref{2}) and (\ref{3}) we have \ba
\nabla \cdot (W^{\xi}\nabla
W)&=&\left.\frac{W^{\xi+1}}{p(p+1)}[\xi\big(W^{-2}\bar{R}-R^{\mu}_{\mu}\big)
\right.\nonumber\\&+&\left.(p-\xi)\big(\tilde{R}-R^{a}_{a}\big)].\right.\label{7}
\ea

By using the $D$-dimensional Einstein equation \be R_{AB}=8\pi
G_{D}\Big(T_{AB}-\frac{1}{D-2}G_{AB}T\Big),\label{8} \ee where
$G_{D}$ is the gravitational constant in $D$ dimensions, it is
easy to write down the following equations \ba
R^{\mu}_{\mu}=\frac{8\pi
G_{D}}{D-2}(T^{\mu}_{\mu}(D-p-3)-T^{m}_{m}(p+1)),\label{9}\\
R^{m}_{m}=\frac{8\pi
G_{D}}{D-2}(T^{m}_{m}(p-1)-T^{\mu}_{\mu}(D-p-1)).\label{10}\ea In
the equations above we set
$T^{\mu}_{\mu}=W^{-2}g_{\mu\nu}T^{\mu\nu}$, so that
$T^{M}_{M}=T^{\mu}_{\mu}+T^{m}_{m}$. Now, it is possible to relate
$R^{\mu}_{\mu}$ and $R^{m}_{m}$ in equation (\ref{7}) in terms of
the stress-tensor. Note also that the left hand side of the
equation (\ref{7}) vanishes upon a line integration over a closed
path along the compact internal space. Hence, taking all that into
account we have \ba && \oint \left.
W^{\xi+1}\Bigg(T^{\mu}_{\mu}[(p-2\xi)(D-p-1)+2\xi]\right.\nonumber\\&+&\left.
T^{m}_{m}p\,(2\xi-p+1)+\frac{D-2}{8\pi
G_{D}}[(p-\xi)\tilde{R}\right.\nonumber\\&+&\left.\xi
\bar{R}W^{-2}]\Bigg)= 0.\right.\label{11} \ea

This last equation provides a one parameter family of consistency
conditions for warped braneworlds in arbitrary dimensions, just
like in \cite{Leblond}. Each choice of $\xi$ lead to a specific
consistency condition. Let us hereon, particularize the analysis
to the five-dimensional bulk case, since it is, indeed, the
current case in variable tension one-brane models \cite{LAS}. In
such codimension one case, we have $D=5$, $p=3$, and
$\tilde{R}=0$. Besides, we shall consider the specific case where
$\xi=-1$, since it eliminates the overall warp factor, resulting
in the most interesting case. With these assumptions, the equation
(\ref{11}) takes a simple form given by \be \oint
(T^{\mu}_{\mu}-4T^{m}_{m})=\frac{\bar{R}}{8\pi G_{5}}\oint
W^{-2}.\label{12}\ee In order to put the consistence conditions in
terms of the brane tensions let us establish the following
stress-tensor ansatz \ba && \left. T_{MN}=-\frac{\Lambda
G_{MN}}{8\pi
G_{5}}+\tau_{MN}\right.\nonumber\\&-&\left.\sum_{i}\Bigg[T_{3}^{(i)}+\kappa^{(i)}\frac{\partial
T_{3}^{(i)}}{\partial t}\Bigg]
P[G_{MN}]_{3}^{(i)}\Delta(r-r_{i}),\right.\label{13} \ea where
$\Lambda$ is the bulk cosmological constant, $\tau_{MN}$ is the
bulk matter fields stress-tensor, $T_{3}^{(i)}$ is the (time
variable) tension of the $i^{th}$ 3-brane, $P[G_{MN}]_{3}^{(i)}$
is the pull-back of the metric to the 3-brane and
$\Delta(r-r_{i})\equiv
\delta(r-r_{i})/\sqrt{G_{rr}}=\delta(r-r_{i})$ stands for the
covariant delta function necessary to position the brane. The time
variation of the $i^{th}$-brane tension is, as usual, computed by
the $\partial_{t}T_{3}^{(i)}$ term and the positive constant
$\kappa^{(i)}$, responsible to the magnitude of the variation, has
units of (energy)$^{-1}$, while the brane tension still has units
of [energy/(lenght)$^{3}$]. We emphasize that the equation
(\ref{13}) is nothing but a slightly modification of the ansatz
presented in \cite{Leblond}. Imposing $\kappa^{(i)}$ to vanish or,
equivalently, taking both the tensions constant, the same
stress-tensor of reference \cite{Leblond} is
obtained\footnote{Compare with equation (14) of reference
\cite{Leblond}.}.

We remark that the generalization implemented by equation
(\ref{13}) is just a first approximation, since it only takes
linear contributions coming from the variation of the tension. We
shall, however, keep our analysis in this straightforward case for
two main reasons: firstly, as we will see, there are important
results coming from such simplest case. Secondly, there are no
physical reasons to encode higher derivatives as well as non
linear terms in the time variation of the tension in the classical
approach. The possibility of a spatial variation of the tension
will be briefly discussed in the final Section.

Before computing the necessary partial traces of the stress-tensor
to complete the analysis via equation (\ref{12}), let us assume
for simplicity $\tau_{MN}=0$, i. e. there is no contribution from
bulk matter fields. Hence, it is easy to see from (\ref{13}) that
\ba T^{\mu}_{\mu}=\frac{-4\Lambda}{8\pi
G_{5}}-4\sum_{i}\Big[T_{3}^{(i)}+\kappa^{(i)}\partial_{t}T_{3}^{(i)}\Big]\Delta(r-r_{i})\label{14}\ea
and \ba T^{m}_{m}=\frac{-\Lambda}{8\pi G_{5}}.\label{15}\ea

Taking into account the equations (\ref{12}), (\ref{14}), and
(\ref{15}), we find \be -\bar{R}\oint W^{-2}=32\pi
G_{5}\sum_{i}\Big[T_{3}^{(i)}+\kappa^{(i)}\partial_{t}T_{3}^{(i)}\Big].\label{16}
\ee Note that in the equation above the bulk cosmological constant
is factored out. Furthermore, the equation (\ref{16}) is valid
only for tensions which do not depend on the extra dimension. Now,
in trying to describe our universe, one can affirm that
$\bar{R}=0$ with an accuracy of $10^{-120}M_{Pl}$, where $M_{Pl}$
is the four-dimensional Planck mass. Then, the final result of the
consistency analysis appears to be quite simple. In fact, from
equation (\ref{16}) one has simply \ba
\sum_{i}\Big[T_{3}^{(i)}+\kappa^{(i)}\partial_{t}T_{3}^{(i)}\Big]=0.
\label{nova1}\ea Note that in the context of constant tensions we
recover the well known RS fine tuning between the brane tensions
\ba T_{3}^{vis}=-T_{3}^{hid}\label{nova2},\ea as expected.

Hereon, some specific cases are analyzed in order to obtain
physical insights about the variable tension possibility. We start
by showing an aggravation of the fine tuning problem, appearing in
the scope of time variable tensions.

\subsection{Eötvös branes and the dynamical fine tuning}

The first case we consider is when both branes are endowed of
tensions obeying the Eötvös law $T_{3}^{(i)}=\pm\lambda^{(i)}t$.
To do so respecting the positivity of the $T_{3}^{vis}$ we
impose\footnote{Note that for our universe in a RS-like model the
brane tension must be positive in order to recover a positive
Newton's
gravitational constant.} \ba T_{3}^{vis}=\lambda^{vis}t \nonumber, \\
T_{3}^{hid}=-\lambda^{hid}t,\label{nova4}\ea where, as remarked,
$\lambda^{(i)}>0$. Substituting the equations (\ref{nova4}) into
(\ref{nova1}) one obtains easily the following constraints \ba
\lambda^{vis}=\lambda^{hid},\label{nova5}\ea and \ba
\kappa^{vis}=\kappa^{hid}.\label{nova6}\ea However, in the light
of equations (\ref{nova4}), it means that the fine tuning must be
time dependent, i. e., \ba T_{3}^{vis}+T_{3}^{hid}=0 \Rightarrow
T_{3}^{vis}(t)+T_{3}^{hid}(t)=0\label{nova7}.\ea Therefore, in
order to satisfy the consistency conditions the time variable
Eötvös brane tensions must obey a dynamical fine tuning. It is
indubitably a worsening of the previous (static tension) case. In
particular, the relation (\ref{nova7}) indicates that both branes
need, at least, to respect some global (unknown) dynamics. It is
hard to accept that the condition (\ref{nova7}) is satisfied by
chance. As a first consequence, we emphasize by passing, a weak
brane tension (as on the hidden brane at later times) complicates
the modulo stabilization mechanism \cite{GW}, indicating once more
the necessity of including the backreaction on the metric to
stabilize the distance between the branes \cite{GKL}.

In order to circumvent the cumbersome situation of a dynamical
fine tuning, several different approaches that lead to consistency
conditions other then (\ref{nova1}) can be used. Here we just
enumerate some possibilities. The first possibility one can try is
by working in a model with more than one extra dimension. In this
case $\tilde{R}$ can be different from zero in (\ref{11}) and the
bulk cosmological constant is not factored out, both contributing
to relax the constraint (\ref{nova1}). In this case two
possibilities emerge: higher codimension scenarios and hybrid
compactification models (when there is codimension one with a
$p$-brane, being $p>3$). If the codimension is greater than one it
is very difficult to extract gravitational information from the
system. There are very interesting results in codimension two
models \cite{RUTHS,TRA}, but it is far from being a closed issue.
Indeed, for higher codimension the problem is completely open.
Another possibility, the hybrid compactification models, is
potentially interesting, specially in the context of
$\mathbb{R}^{10}\times S^{1}/\mathbb{Z}_{2}$ Horava-Witten
compactifications \cite{HW}. But there are strong restrictions,
coming from experiment, concerning to the size of the extra
on-brane dimensions \cite{HOYLE}. Such constraints can, in some
cases, aggravate the hierarchy problem for static brane tension
models. To sum up, more dimensions do not seem to be the best
strategy to solve the dynamical fine tuning problem.

In the five-dimensional General Relativity scope, one can scape
from the implications of equations (\ref{nova1}) and (\ref{nova4})
by allowing the presence of (unknown) matter fields in the bulk.
Obviously, if $\tau_{MN}$ is not zero in (\ref{13}) the sum over
the branes tensions (plus time variations) can be counterbalanced
by the presence of the traces ($\tau^{m}_{m},\tau^{\mu}_{\mu}$) in
the consistency conditions. Of course, it can lead to another
strong constraint, this time on the bulk matter but the dynamical
fine tuning is not necessary anymore. The crucial problem arising
from this approach rests, perhaps, on the fact that we do not have
any information {\it a priori} about the behavior, and even the
nature, of those bulk fields.

In this vein, we shall point out another way to overcome the
dynamical fine tuning problem simply by lifting the necessity of
two Eötvös branes tensions at the same time. In fact, in the light
of the previous discussion, two Eötvös branes appear to be
problematic enough. So, the first different situation that might
be thought is when we impose that, just one of the two mirror
branes obeys the Eötvös law. More precisely, what happens if we
fix the behavior of the hidden brane as an Eötvös brane? Before
looking at that point, let us briefly comment one more situation
in the next Section.

\subsection{Independent branes tensions}

The equation (\ref{nova1}) means that the sum of the bracket terms
must vanish. However it is possible that the terms
$T_{3}^{(i)}(t)+\kappa^{(i)}\partial_{t}T_{3}^{(i)}(t)$ vanish for
each brane independently. This is not the most interesting case,
nevertheless we shall make a brief comment on it, since it will be
important for the next case.

Setting $T_{3}^{(i)}(t)+\kappa^{(i)}\partial_{t}T_{3}^{(i)}(t)=0$
for all $(i)$ one finds \ba T_{3}^{(i)}(t)\sim
\exp{(-t/\kappa^{(i)})}.\label{nova8} \ea So, in a scenario where
the branes do not feel any influence from each other, the shape of
the tension time variation must be like the one given by equation
(\ref{nova8}). Note that the exponential behavior of
$T_{3}^{(i)}(t)$ is dominated by the constant $\kappa^{(i)}$, i.
e., for a weak (strong) variation coupling, $\kappa^{(i)}\gg 1$
$(\ll 1)$, the tensions time variation is also weak (strong). Now
let us analyze the behavior of the visible brane tension when the
hidden brane tension obeys an Eötvös law.

\subsection{Bouncing-like behavior}

A very interesting case is found when one of the branes tension
configuration is taken into account, with the constraint that the
another brane tension must obey the Eötvös law. In order to
implement it, let us impose $T_{3}^{hid}=-\lambda^{hid}t$.

Substituting the previous requirement into the consistency
condition (\ref{nova1}) we arrive at the following ordinary
differential equation \ba
\frac{dT_{3}^{vis}(t)}{dt}+\frac{T_{3}^{vis}(t)}{\kappa^{vis}}-\frac{\lambda^{hid}}{\kappa^{vis}}t-
\frac{\kappa^{hid}\lambda^{hid}}{\kappa^{vis}}=0,\label{nova9} \ea
whose solution is given by \ba
T_{3}^{vis}(t)=\lambda^{hid}t+Ce^{-t/\kappa^{vis}}+\lambda^{hid}\Big(\kappa^{hid}-\kappa^{vis}\Big),\label{nova10}
\ea where $C$ is an arbitrary integration constant with
(energy)$^{-2}$ units, just as $\lambda^{(i)}t$. Absorbing the
last factor into the visible brane tension, equation
(\ref{nova10}) reads \ba
T_{3}^{vis}(t)=-T_{3}^{hid}(t)+Ce^{-t/\kappa^{vis}}.\label{nova11}
\ea The first characteristic to note is the mixed behavior of
$T_{3}^{vis}$. It is governed by a negative exponential at early
times in such way that the tension was decreasing. Nevertheless
after a critical time, say $\bar{t}$, the dominant term is given
by the Eötvös law of the hidden brane. It is easy to compute the
critical time $\bar{t}$, since it is the minimal point of the
$T_{3}^{vis}$ function. By imposing $\frac{dT_{3}^{vis}(t)}{dt}=0$
it follows that \ba
\bar{t}=\kappa^{vis}\ln\Bigg(\frac{C}{\lambda^{hid}\kappa^{vis}}\Bigg)\label{nova12},
\ea in such way that the constant $C$ should be positive. So, it
is reasonable to interpret $C$ as an independent contribution to
$T^{(vis)}$.

\begin{figure}[!h]
  \includegraphics[width=6cm]{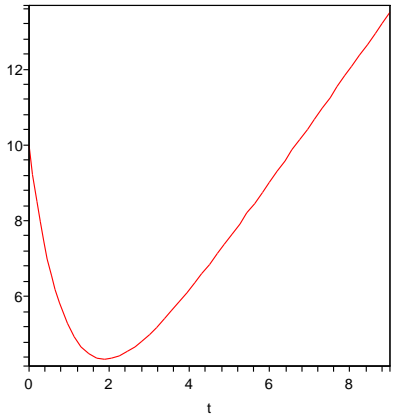}\\
  \caption{Illustrative graphic of $T_{3}^{vis}(t)$ for numerical values $\lambda^{hid}=1.5$, $\kappa^{vis}=1$, and $C=10$.
  Note the bouncing-like behavior before the minimal point $(\bar{t}\approx 1.897)$. A fine tuning between
  $\lambda^{hid}$, $\kappa^{vis}$, and $C$ $(C=\lambda^{hid}\kappa^{vis})$ eliminates the bouncing-like period.}\label{Figure 1}
\end{figure}

The qualitative behavior of $T_{3}^{vis}(t)$ can be better
visualized in the Figure 1. The usual interpretation of this
result is given as follows: for times $t>\bar{t}$ the exponential
term is suppressed and the dominant term of the visible brane
tension is given by the Eötvös one. As time goes the brane becomes
more rigid, and the universe expands. In particular this dominant
term is given by the hidden brane tension with opposite signal.
For times $t<\bar{t}$, however, the dominant factor is the
exponential one. Nevertheless, as time flows the visible brane
tension decreases allowing a bouncing-like period for the universe
described by such brane. It is an interesting characteristic of
this toy model. It emerges naturally from the consistency
conditions in five dimensions by requiring an Eötvös hidden brane
tension. We should emphasize that this bouncing-like period can be
eliminated by imposing the additional fine tuning on the couplings
\ba C=\lambda^{hid}\kappa^{vis},\label{nova13}\ea as one can see
from equation (\ref{nova12}).

The functional form of the visible brane tension can be used then,
at least qualitatively, as a tool to understand the general
behavior of the universe it describes. Returning again to the
content of the Figure 1, one sees that at early times
$(0<t<\bar{t})$ the visible brane has, predominantly, a
independent behavior (in the sense of the previous subsection)
from the hidden brane. It contracts until $\bar{t}$, and after
that the effects coming from the presence of the hidden brane
begin to dominate.

The complete dynamics involved in such a process is far beyond the
scope of this paper. The point, however, is that some interesting
possibilities arise even studying simple models like that, based
upon the formal context of the consistency conditions.

\section{Concluding Remarks and Outlook}

By applying the consistency conditions to five-dimensional
variable tension branes in a RS-like setup we investigate the
behavior of some time variable tension specific scenarios. In
particular we demonstrate that the requirement of two Eötvös
branes seems to aggravate the fine tuning problem, indicating that
both branes should obey a more general unknown dynamic.

An interesting scenario arises, however, when we require that just
one of the branes respects the Eötvös law. For instance, if the
hidden brane satisfies this last requirement, the visible brane
tension shows up a mixed behavior indicating a bouncing-like
period at early times. It occurs when the visible brane seems not
to ``feel'' the hidden brane, after which an expansion period,
encoded in a mainly Eötvös tension term, comes to dominate. We
remark that the contracting period can be eliminated by an
additional fine tuning in the couplings of the model.

This work deals only with the possibility of time variable brane
tensions, in particular when time variations related by the Eötvös
law, since it is a physically motivated one \cite{LAS}. We shall
however make some brief comments about the possibility of a
spatial tension variation. Perhaps a more complete scenario can
arise from the implementation of a spatial variation in the
tension in addition to the time variation. In the context of
consistency conditions, it can be accomplished by another
extension of the stress-tensor (\ref{13}). We expect a class of
non-trivial solutions arising from such an extension, since the
spacial variations must be suppressed by, for instance, a time
damping factor in order to reproduce the observable isotropic
universe at large scales. However, a definitive word requires a
more careful analysis.

After these brief notes concerning the viability of variable
tensions braneworld models, we shall use part of this final
Section pointing out some possible future research lines, in order
to bring some reasons why the analysis of such scenarios are
worthwhile. Firstly, it has to be taken very seriously how to
match an essentially anisotropic brane coming from some variable
tension models with a large scale isotropic universe, as the one
we experiment. As showed in \cite{LAS} it is not the case of
Eötvös branes in the one-brane setup. However, new cosmological
signatures are also expected in the scope of those (not-Eötvös)
variable tension branes. The anisotropic background generated by
such branes in the two-brane models breaks some standard model
fields symmetries as well as can lead to subtle violation of the
Equivalence Principle. Therefore, it constitutes a good laboratory
to the study of fundamental principles limits.

Apart of that, a variable tension braneworld model can also
incorporate, at least in some regime, the production of branons
\cite{prebranons,branons}. Starting from the fact that any brane
must fluctuate, because a completely rigid object cannot exist in
the scope of a relativistic theory, one concludes that a new
(scalar) field has to be included to represent the brane position
in the bulk. Such field gives rise to Goldstone bosons (the
branons) arising from the spontaneous symmetry breaking of the
full bulk diffeomorphism due the presence of the brane. It seems
that when the brane is flexible enough (with a tension scale much
smaller then the fundamental gravity scale), branons are the
unique particles to be expected. We remark that this kind of
analysis was also developed in the context of orbifold
compactifications \cite{orbif}. In the context analyzed in the
previous Section it is not ruled out a set of parameters for
branons production on the visible brane, for instance, in the
early universe.

We conclude by saying that variable tension braneworlds can also
impose an imprint in effective projected four-dimensional field
theory, since the Kaluza-Klein mass spectrum of higher dimensional
theories is indirectly affected by the brane tension. Therefore,
it can, potentially, lead to new phenomena in particle physics. We
shall finish emphasizing the usefulness of the consistency
conditions in the establishment of this type of braneworld models,
in the sense of avoid ill defined scenarios. In this vein, the
consistency conditions are also useful as starting point of
further investigations in variable tension models, in order to
study links with experimental signatures coming from new
phenomenology obtained by studying mathematically consistent
models.

\section*{Acknowledgments}

The authors thanks to prof. M. E. X. Guimarães for useful
conversation and prof. Ion Vancea for many interesting viewpoints,
discussions, and for calling our attention for such a variety of
references. We also thanks to an anonymous referee for important
comments about the paper. M. C. B. Abdalla acknowledges CNPq for
support, J. M. Hoff da Silva thanks to CAPES-Brazil for financial
support.

\end{document}